\documentclass[aps,amsfonts,amsmath,amssymb,pra,showpacs]{revtex4}

\usepackage{graphicx}
\newcommand{\be}{\begin{equation}}
\newcommand{\ee}{\end{equation}}
\newcommand{\bex}{\begin{eqnarray}}
\newcommand{\eex}{\end{eqnarray}}
\newcommand{\bsub}{\begin{subequations}}
\newcommand{\ensub}{\end{subequations}}
\newtheorem{thm}{Theorem}

\def\qed{$\blacksquare$}

\newcommand{\ou}{\rm ~OR~}
\newcommand{\og}{\rm ~AND~}
\newcommand{\ouh}{\rm ~OR}

\begin{document}
\title{Some Directions beyond Traditional Quantum Secret Sharing} 

\author{Sudhir Kumar Singh}
\email{suds@ee.ucla.edu}
\affiliation{Department of Electrical Engineering, University of California, Los 
Angeles, CA 90095.} 
\author{R.   Srikanth}    
\email{srik@rri.res.in}
\affiliation{Poornaprajna Institute of Scientific Research,
Devanahalli, Bangalore- 562 110, India.}
\affiliation{Raman  Research   Institute,  
Sadashiva nagar, Bangalore-560080, India}

\begin{abstract}
We  investigate two  directions  beyond the  traditional  quantum secret  sharing (QSS). First,  a
restriction  on QSS that  comes from  the
no-cloning theorem  is that any pair  of authorized sets  in an access
structure  should overlap.   From the  viewpoint of  application, this
places  an  unnatural constraint  on  secret  sharing.   We present  a
generalization,  called assisted QSS  (AQSS), where  access structures
without pairwise  overlap of authorized sets  is permissible, provided
some shares  are withheld by the  share dealer.  We show  that no more
than $\lambda-1$ withheld shares  are required, where $\lambda$ is the
minimum number of {\em  partially linked classes} among the authorized
sets  for the  QSS. Our
result means that such applications of QSS need not be thwarted by the
no-cloning theorem.  Secondly, we  point out a  way of  combining the
features of  QSS and quantum  key distribution (QKD)  for applications
where a classical  information is shared by quantum  means. We observe
that  in such  case, it  is often possible  to reduce  the security
proof of QSS to that of QKD.

\end{abstract}

\maketitle

\section{Introduction\label{intro}}

Suppose the  president of  a bank,  Alice, wants to  give access  to a
vault  to two  vice-presidents, Bob  and  Charlie, whom  she does  not
entirely trust.  Instead of giving the combination to any one of them,
she may  desire to distribute  the information in  such a way  that no
vice-president alone has any knowledge of the combination, but both of
them can jointly determine the combination.  Cryptography provides the
answer  to  this  question  in   the  form  of  {\it  secret  sharing}
\cite{schneier96}.  In this scheme, some sensitive data is distributed
among a number of parties such that certain authorized sets of parties
can  access  the  data,  but  no  other  combination  of  players.   A
particularly symmetric variety of secret splitting (sharing) is called
a  {\it threshold  scheme}: in  a $(k,n)$  classical  threshold scheme
(CTS), the secret  is split up into $n$ pieces  (shares), of which any
$k$ shares form a set  authorized to reconstruct the secret, while any
set  of $k-1$ or  fewer shares  has no  information about  the secret.
Blakely  \cite{blakely79} and  Shamir \cite{sha79}  showed  that CTS's
exist for all values of $k$  and $n$ with $n \geq k$. By concatenating
threshold  schemes,  one can  construct  arbitrary access  structures,
subject only  to the condition  of monotonicity (ie.,  sets containing
authorized sets should also  be authorized) \cite{ben90}. Hillery {\em
et al.}  \cite{hil00} and  Karlsson {\em et al.} \cite{kar00} proposed
methods for  implementing CTSs that  use {\em quantum}  information to
transmit shares securely in the presence of eavesdroppers.

Subsequently,  extending the above  idea to  the quantum  case, Cleve,
Gottesman  and Lo \cite{cle00},  using the  notion of  quantum erasure
correction  \cite{cs,gra97,nc00}, presented  a  $(k,n)$ {\it  quantum}
threshold  scheme (QTS)  as a  method to  split up  an  unknown secret
quantum  state   $|S\rangle$  into   $n$  pieces  (shares)   with  the
restriction that  $k > n/2$--  this inequality being needed  to ensure
that no two disjoint sets of players should be able to reconstruct the
secret,   in   conformance  with   the   quantum  no-cloning   theorem
\cite{woo82}.   QSS has  been extended  beyond QTS  to  general access
structures \cite{got00,smi00},  but here  none of the  authorized sets
shall be  mutually disjoint:  given a QSS  access structure  $\Gamma =
\{\alpha_1,\cdots,\alpha_r\}$   over  $N$   players,   the  no-cloning
restriction entails that:
\begin{equation}
\label{noklo}
\alpha_j \cap \alpha_k \ne \phi~~~~~\forall j,k.
\end{equation}
Potential applications of QSS include creating joint checking accounts
containing quantum  money \cite{wiesner83}, or  sharing hard-to-create
ancilla states \cite{got00},  or performing secure distributed quantum
computation \cite{cre01}.   A tri-qubit  QSS scheme has  recently been
implemented \cite{lan04}.  The  chances of practical implementation of
QSS  are improved by  employing equivalent  schemes that  maximize the
proportion of classical information processing \cite{nas01,sud03}.
It has been shown that quantum teleportation \cite{bang} and 
entanglement swapping \cite{cab00,kar02} may be used to implement
an $((n,n))$-threshold scheme.

The requirement Eq. (\ref{noklo}) places a restriction quite unnatural
to applications,  where we  may more likely  expect to find  groups of
people with mutual trust within  the group, and hardly any outside it.
First aspect of our  present  work  is  aimed  at  studying a  way  to  
overcome  this
limitation.   In  particular, in the Sections \ref{sec:plc} and \ref{sec:qencry} 
we  show  that  allowing  the dealer  to
withhold a small number  of shares permits arbitrary access structures
to be acceptable, subject  only to monotonicity.  This modified scheme
we  call ``assisted  QSS'' (AQSS),  the shares  withheld by  the dealer
being called ``home shares''. While more general than conventional QSS,
AQSS is clearly  not as general as classical  secret sharing, since it
requires  shares given  to the  (non-dealer) players,  called ``player
shares'', to  be combined with  the home shares for  reconstructing the
secret.

Inspite of this limitation, the  modified scheme can be useful in some
applications  of secret  sharing, in  particular, those  in  which the
secret   dealer  is   by  definition   a  trusted   party   and  where
reconstruction of  the secret effectively occurs  by re-convergence of
shares at  the dealer's  station. Further, it could be  useful for  schemes like
circular QSS \cite{cqss}. 

We note that the home share by themselves give no information.  In the
bank example above, access is allowed  by the bank vault (which can be
thought of effectively  as the dealer, acting as  the bank president's
proxy) if the secret reconstructed from the vice-presidents' shares is
the required password. The locker  thus effectively serves as both the
dealer and site of secret  reconstruction.  In AQSS, the player shares
are   combined  with   the  home   share(s)  to   reconstruct  the
secret. Clearly, this  leads to no loss of generality  in this type of
QSS.  Where the  secret dealer is not necessarily  trusted, such as in
multi-party secure computation (MPSC), AQSS may be less useful, though
here again only a more detailed  study can tell whether MPSC cannot be
turned into a suitable variant of AQSS.

It is  assumed that all the  $n$ (quantum) shares  are somehow divided
among the $N$  players. In an AQSS scheme, $m <  n$ shares are allowed
to remain with  the share dealer, as home shares.   In order that AQSS
should depart minimally from conventional QSS, we further require that
the number  of home shares should  be {\em the  minimum possible} such
that a  violation of  Eq. (\ref{noklo}) can  be accomodated.   Thus, a
conventional QSS  access structure like $\Gamma =  \{ABC, ADE, BDF\}$,
which as such conforms to the no-cloning theorem, will require no {\em
share assistance}.   A conventional  QSS scheme is  a special  case of
AQSS, in  which the set of home  shares is empty.  We  prove by direct
construction that, by  allowing for non-zero
home  shares, the restriction  (\ref{noklo}) does  not apply  to AQSS.
Therefore, with  share assistance, the only restriction  on the access
structure $\Gamma$  in AQSS is monotonicity, as  with classical secret
sharing. Such constructions are described in details in the Sections 
\ref{sec:plc} and \ref{sec:qencry}.

Another cryptographic primitive where (multipartite) entanglement
can be effectively used is that of quantum key distribution (QKD)
\cite{BBqkd,lochau} and its generalization to $n$ parties ($n$-QKD).
Note that the $n$-QKD involves  sharing a random
key  amongst  $n$  {\it trustworthy}  parties  unlike  QSS which splits  quantum
information  among {\it untrustworthy }  parties.  
Naturally, it  would be  an interesting
extension to consider the situations where some kind of  mutual trust may
be present  between sets of  parties while parties being individually
mistrustful wherein it might be possible to  combine the essential
features  of  QKD and  QSS.   In the  Section \ref{sec:kombi},  
we  discuss one  such extension. 
We  consider  the  problem  of  secure key  distribution  between  two
trustful   groups   where   the   invidual  group   members   may   be
mistrustful. The  two groups retrieve  the secure key string,  only if
all members should cooperate with  one another in each group. That is,
how the  two groups one of  size $k$ and  the other of size  $n-k$ may
share an  identical secret key  among themselves while  an evesdropper
may co-operate with several (of  course not all) dishonest members from
any of the groups. If $k=1$, the result is equivalent to a $(n-1,n-1)$
threshold secret sharing scheme.

The  QKD-QSS   connection  is   manifest  in  several   earlier  works
\cite{hil00,bag03,cqss}. Ref. \cite{hil00} first introduced the idea
of using  a GHZ  to implement a  three-party secret  sharing protocol.
Ref. \cite{bag03}  extend a method  of QKD with  reusable entanglement
\cite{zha} to QSS. Deng et  al. \cite{cqss} extend the ping-pong QKD
protocol \cite{bostrom} to a three-party circular QSS scheme.
Ref. \cite{sen03}  presents a security for a  $(n,n)$ scheme involving
on  $n$-partite entangled  states, based  on the  violation  of Bell's
inequalities,  even  when the  $n$-qubit  correlations  are weak.   In
contrast,  we  employ only  bipartite  states.   From the  theoretical
perspective, this  will provide the  simplicity that we can  build the
our protocol on top of QKD,  which will help us reduce the security of
our scheme to that of  QKD. From a practical perspective, multipartite
states employed for QSS above has exponentially low efficiency even in
the  noiseless scheme,  since only  rounds in  which  all participants
measure the  same observable, $\sigma_x$ or  $\sigma_y$, are retained,
with all  other measurement  possibilities discarded. In  contrast, in
our protocol,  the key generation  step will involve a  measurement by
all parties in the diagonal basis,  so that no waste bits are produced
through incompatible measurements by the various parties.

\section{Partial link classification \label{sec:plc}}

Given  access structure  $\Gamma  = \{\alpha_1,\cdots,\alpha_r\}$,  we
divide  all  authorized sets  $\alpha_j$  into  {\em partially  linked
classes},  each  of  which  is  characterized  by  the  following  two
properties: (a) Eq. (\ref{noklo}) is satisfied if $\alpha_j, \alpha_k$
belong to the  same class; (b) for any two  distinct classes, there is
at least  one pair  $j,k$, where $\alpha_j$  belongs to one  class and
$\alpha_k$ to the other, such that Eq. (\ref{noklo}) fails.

A division  of $\Gamma$ into  such classes we  call as a  {\em partial
link  classification}.   The  number  of  classes in  a  partial  link
classification gives  its size.  In general,  neither the combinations
nor size  of partial  link classifications are  unique. We  denote the
size of  the {\em  smallest} partial link  classification for  a given
$\Gamma$  by $\lambda$. If  all authorized  sets have  mutual pairwise
overlap  then $\lambda=1$ and  the single  partially linked  class is,
uniquely, $\Gamma$  itself, and AQSS  reduces to conventional  QSS. If
none  of   the  $\alpha_j$'s   have  mutual  pairwise   overlap,  then
$\lambda=r$ and  the $r$ partially linked classes  are, uniquely, each
$\alpha_j$.   If there  are  $s$ disconnected  groups of  $\alpha_j$'s
(that  is, Eq. (\ref{noklo})  fails for  all pairs  $j, k$,  where $j$
comes from one  group and $k$ from another) then  $\lambda \ge s$. The
inequality  arises from  the  fact that  there  may be  more than  one
partially linked class within a disconnected group.

\begin{figure}
\includegraphics{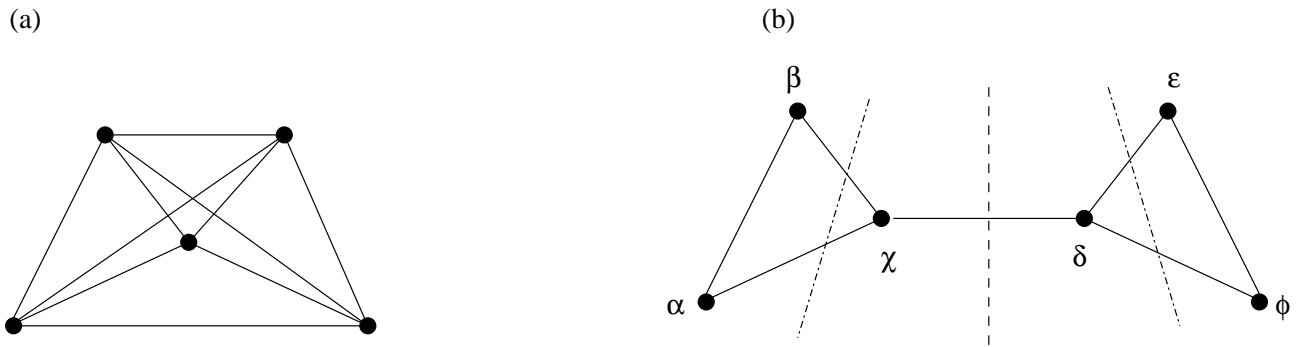}
\caption{The vertices  represent authorized  sets, the edges  depict a
non-vanishing pairwise overlap between two authorized sets. Figure (a)
represents a  conventional QSS, where all sets  have pairwise overlap,
meaning that the AS graph is complete, so that $\lambda=1$. Figure (b)
represents a situation  where this does not hold  and hence $\lambda >
1$. The authorized sets are labelled $\{\alpha, \beta, \chi, \epsilon,
\delta, \phi\}$.  The dashed  line is a  cut leading to  two partially
linked  classes (the  pair of  3-cliques,  $\{\alpha,\beta,\chi\}$ and
$\{\delta,\epsilon,\phi\}$), so that  $\lambda=2$. The two dash-dotted
lines are system of two cuts leading to three partially linked classes
(the  triple of  2-cliques, $\{\alpha,\beta\}$,  $\{\chi,\delta\}$ and
$\{\epsilon,\phi\}$).}
\label{AQSS}
\end{figure}

The problem of obtaining a partial link classification can be analyzed
graph  theoretically.  It  is easy  to visualize  $\Gamma$ as  a graph
$G(V,E)$, composed of a set $V$  of vertices and set $E$ of edges. The
vertices  are the authorized  sets, $V=\{\alpha_j\}=\Gamma$  and edges
$E=\{(\alpha_j,\alpha_k)\}$  correspond  to pairs  of  sets that  have
pairwise overlap.  Such a graph may be called an {\em access structure
graph}  (AS  graph)  for  $\Gamma$.   A  partial  link  classification
corresponds  to a  partitioning of  the AS  graph $G$  such  that each
partition is a {\em clique}, i.e., a {\em complete} subgraph in $G$ (A
graph is called complete if its each vertex has an edge with its every
other vertex).  Figure \ref{AQSS}(a) depicts a conventional QSS, where
$\Gamma$ is  partitioned into a single  5-clique. Figure \ref{AQSS}(b)
depicts  a  more general  case  covered  by  AQSS, where  $\Gamma$  is
partitioned into  a pair of 3-cliques  or into a  triple of 2-cliques.
The  problem  of  determining  $\lambda$  is thus  equivalent  to  the
combinatorial problem  of partitioning $G$ into the  minimum number of
cliques. Here  it is worth  noting that many multi-party  problems are
amenable to combinatorial treatment.

Before introducing the  main result, it is instructive  to look at the
classical  situation.  In  our notation,  single  (double) parantheses
indicate CTS  (QTS).  For a  classical secret sharing  scheme, suppose
$\Gamma = \{ABC,  AD, DEF\}$, which can be written  in the normal form
$\{(A \og B \og C) \ou (A \og  D) \ou (D \og E \og F)\}$. The \og gate
corresponds to a $(|\alpha_j|,|\alpha_j|)$ threshold scheme, while \ou
to a (1,2) threshold scheme. By concatenating these two layers, we get
a construction for $\Gamma$.  In the conventional QSS, the above fails
for two reasons, both connected to the no-cloning theorem: the members
of $\Gamma$ should not be  disjoint; and further there is no $((1,2))$
scheme.   

However,  we  can  replace  $((1,2))$  by a  $((2,3))$  scheme,  which
corresponds to a majority function  of \ouh.  In general, we replace a
$((1,r))$  scheme  by  a  $((r,2r-1))$  scheme.   $r$  of  the  shares
correspond to individual authorized sets in $\Gamma$, shared within an
$\alpha_j$   according  to  a   $((|\alpha_j|,|\alpha_j|))$  threshold
scheme, and, recursively, the  other $r-1$ shares are shared according
to  a   pure  state  scheme   that  implements  a   maximal  structure
$\Gamma_{\max}$ that includes  $\Gamma$ (obtained by adding authorized
sets to  $\Gamma$ until  the complement of  every unauthorized  set is
precisely  an  authorized  set) \cite{got00}.  Theorem  \ref{thm:gams}
below  extends this  idea to  the  situation where  $\Gamma$ does  not
satisfy Eq. (\ref{noklo}).
\begin{thm}
\label{thm:gams}
Given   an   access   structure   $\Gamma  =   \{\alpha_1,   \alpha_2,
\cdots,\alpha_r\}$  with  a  minimum  of  $\lambda$  partially  linked
classes among a set of players ${\cal P} = \{P_1, P_2, \cdots, P_N\}$,
an  assisted quantum  secret  sharing scheme  exists  iff $\Gamma$  is
monotone. It requires no more than $\lambda-1$ home shares.
\end{thm}
\noindent
{\bf  Proof.} It is  known that  if $\lambda=1$,  then there  exists a
conventional QSS to realize  it \cite{got00}.  Suppose $\lambda>1$. To
implement  $\Gamma$ (which represents  a monotonic  access structure),
the dealer first employs a $((\lambda,2\lambda-1))$ majority function,
assigning one  share to each  class.  Recursively, each share  is then
subjected  to a  conventional  QSS within  each  class. The  remaining
$\lambda-1$  shares  remain  as   home  shares  with  the  dealer.  To
reconstruct the  secret, any authorized set can  reconstruct the share
assigned  to its  class,  which,  combined with  the  home shares,  is
sufficient for the purpose.  Clearly,  since the necessity of the home
share by  itself fulfils the  no-cloning theorem, authorized  sets are
{\em not}  required to be  mutually overlapping. Thus  monotonicity is
the only constraint. \hfill \qed
\bigskip

Some corrolories  of the theorem are  worth noting. First  is that the
number ($=\lambda-1$) of home shares  is strictly less than the number
($\ge N \ge \lambda$) of player shares.  A share $q$ is `important' if
there  is  an  unauthorized  set  $T$  such that  $T  \cup  \{q\}$  is
authorized. From  the fact  the Theorem uses  a threshold  scheme (the
$((\lambda,2\lambda-1))$ scheme)  in the first layer,  it follows that
all the home shares are important.

As an  illustration of Theorem \ref{thm:gams}, we  consider the access
structure $\Gamma = \{ABC, BD,  EFG\}$, for which $\lambda=2$.  In the
first layer, a $((2,3))$ scheme  is employed to split $|S\rangle$ into
three  shares, with  one share  designated  to the  class $C_1  \equiv
\{ABC, BD\}$ and  the other to $C_2 \equiv  \{EFG\}$. The last remains
with the dealer.  In the second layer, the first share is split-shared
among members  of $C_1$ according  to a conventional QSS  scheme.  The
second share  is split-shared  among players of  $C_2$ according  to a
$((3,3))$ scheme. Diagrammatically, this can be depicted as follows.
\begin{equation}
\label{eq:qz0}
((2,3)) \left\{ \begin{array}{ll}
((2,3)) : & \left\{ \begin{array}{ll} 
				((3,3)) : & A, B, C \\
				((2,2)) : & B, D    \\
				|S^{\prime}\rangle
		    \end{array} \right. \\
 ~~ & ~~ \\
((3,3)) : & E, F, G \\
 ~~ & ~~ \\
((1,1)) : &  {\rm dealer}
	   \end{array} \right. 
\end{equation}

Note  that given any  $\Gamma$, even  with disconnected  elements, (so
that the AS graph is not connected), there is an AQSS by simply adding
a common player to all authorized  sets, and designating him to be the
dealer:  eg., $\Gamma =  \{ABC, DE,  FGH\}$ giving  $\Gamma^{\prime} =
\{ABCX, DEX, FGHX\}$, where shares  to $X$ would be designated as home
shares.        Thereby,       the       structure      $\Gamma       =
\Gamma^{\prime}|_{\overline{X}}$,  which   denotes  a  restriction  of
$\Gamma^{\prime}$ to  members other than $X$,  is effectively realized
among the other players. However, this is not an efficient AQSS scheme
because the number of resultant  home shares are non-minimal, at least
according to  the recursive  scheme outlined above.   In all  it would
require $3 + 2x$ shares, where $x$ is the number of instances in which
$X$  appears  in  a  maximal structure  $\Gamma^{\prime}_{\max}$  that
includes  $\Gamma^{\prime}$.   More generally,  the  requirement is  a
minimum  of $r  + (r-1)x$  home  shares, where  $r$ is  the number  of
authorized sets  in $\Gamma$.   A better method  is for the  dealer to
employ a  pure state scheme  that implements $\Gamma^{\prime}_{\max}$,
retain  all shares  corresponding to  $X$ while  discarding  all those
corresponding to sets in $\Gamma^{\prime}_{\max}-\Gamma^{\prime}$.  In
all this would require $3$  shares, or, in general, $r$ shares. Better
still, according to Theorem \ref{thm:gams}, no more than $\lambda-1=2$
home shares are needed. Clearly, in general, $\lambda - 1 < r$.  These
considerations suggest that  $\lambda(\Gamma)-1$ is the minimal number
of home shares required to implement a QSS for $\Gamma$. We conjecture
that this is indeed the case.

\section{AQSS with quantum encryption\label{sec:qencry}}
Finally, we note  that practical AQSS can be  made highly efficient in
terms  of using  quantum  resources by  employing quantum  encryption.
Indeed,  it is  quite useful  in QSS  even outside  the  AQSS paradigm
\cite{nas01,sud03}.  Quantum encryption works  as follows:  suppose we
have a  $n$-qubit quantum state  $|\psi\rangle$ and a  random sequence
$K$ of $2n$ classical bits.   Each sequential pair of classical bit is
associated   with  a   qubit  and   determines   which  transformation
$\hat{\sigma}   \in   {I   ,   \hat{\sigma}_x   ,   \hat{\sigma}_y   ,
\hat{\sigma}_z }$ is applied to  the respective qubit.  If the pair is
00, I is applied, if it is 01, $\hat{\sigma}_x$ is applied, and so on.
To  one not  knowing $K$,  the resulting  $|\tilde{\psi}\rangle$  is a
complete mixture and no information can be extracted out of it because
the encryption leaves any pure  state in a maximally mixed state, that
is:         $(1/4)(\hat{I}|S\rangle\langle         S|\hat{I}         +
\hat{\sigma}_x|S\rangle\langle            S|\hat{\sigma}_x           +
\hat{\sigma}_y|S\rangle\langle            S|\hat{\sigma}_y           +
\hat{\sigma}_z|S\rangle\langle S|\hat{\sigma}_z)  = (1/2)\hat{I}$, for
any  $|S\rangle$.  However,  with  knowledge of  $K$  the sequence  of
operations  can be reversed  and $|\psi\rangle$  recovered. Therefore,
classical data can be used to encrypt quantum data.

In general, given $d$-dimensional objects, quantum encryption requires
$d^2$ operators and  a key of $2\log(d)$ bits  per object to randomize
perfectly \cite{mos00}. In practice, such  quantum operations may prove costly, and
only  near-perfect security  may be  sufficient. In  this  case, there
exists  a set of  roughly $d\log(d)$  unitary operators  whose average
effect on every  input pure state is almost  perfectly randomizing, so
that the  size of  the key  can be reduced  by about  a factor  of two
\cite{haydn}.

The idea  is quite  simple: to share  a quantum  secret $|\psi\rangle$
according  to access  structure $\Gamma$,  which does  not necessarily
satisfy the no-cloning condition,  the dealer first encrypts the state
to  $\tilde{\psi}\rangle$  using classical  bit  string  $K$.  In  the
extreme  case, the  entire encrypted  quantum  state is  treated as  a
single home share and transmitted  to the reconstructor. String $K$ is
then  classically shared  according to  $\Gamma$.  Any  authorized set
$\alpha$  can  reconstruct  $K$,   and  thus  $|\psi\rangle$,  at  the
reconstructor's location.

The above AQSS scheme with quantum encryption is much simpler than 
that based on partial linked classes proposed in Section \ref{sec:plc},
however it is interesting to note that the later scheme is stronger 
in the sense that the  dealer (reconstructor) can give 
his all shares to some third party which might be untrustworthy 
and the secret still remains hidden even if all {\it classical} 
information leaks. Giving away the shares to third party is 
good from practical point of view as the dealer (reconstructor) might be limited 
by quantum memory requirements.  

\section{Combining QKD and QSS\label{sec:kombi}}

In this section, we discuss our protocol for the {\it two-group QKD} - 
the  problem  of  secure key  distribution  between  two
trustful   groups   where   the   invidual  group   members   may   be
mistrustful. The  two groups retrieve  the secure key string,  only if
all members should cooperate with  one another in each group. That is,
how the  two groups one of  size $k$ and  the other of size  $n-k$ may
share an  identical secret key  among themselves while  an evesdropper
may co-operate with several(of  course not all) dishonest members from
any of the groups. If $k=1$, the result is equivalent to a $(n-1,n-1)$
threshold secret sharing scheme.

Note that the two-group QKD  is trivially  a classical secret  sharing scheme  if we
involve a trusted third party, say, Lucy.  Lucy will simply generate a
random classical bit string. Since  it is just a classical information,
she  makes two  copies  of it  and  splits one  each  amongst the  two
groups. Principles  of quantum  physics allows us,  as in the  case of
2-QKD \cite{BBqkd,lochau}, to  do away with  the third party.   
We observe that  the above
problem essentially seems  to be a combination of  (a) $2$-QKD between
the two groups, each group being  considered as a single party and (b)
secret sharing in each group among their parties.

Our  protocol  works  in two  broad  steps.  In  the first  step,  the
$n$-partite problem  is reduced to a  two-party problem by  means of a
method for  teleporting entanglement \cite{skp}.  This  creates a pure
$n$-partite maximally entangled state among $n$-parties, starting from
$n-1$ EPR  pairs shared  along a spanning  EPR tree, using  only ${\cal
O}(n)$   bits   of   classical   communication.    This   entanglement
teleportation protocol  exploits the combinatorial  arrangement of EPR
pairs to simplify the  task of distributing multipartite entanglement.
In  the second  step,  as in  case  of $n$-QKD,  the Lo-Chau  protocol
\cite{lochau} or  Modified Lo-Chau protocol \cite{sp00}  is invoked to
prove the  unconditional security of sharing nearly  perfect EPR pairs
between two parties.

To this end, we will be using a state of the form:
\begin{equation}
\label{maxeq}
|\Psi\rangle = \frac{1}{\sqrt{2}}(|00\cdots 0\rangle + |11\cdots1\rangle),
\end{equation}
a   maximally  entangled   $n$-partite  state,   represented   in  the
computational basis.

Our protocol is motivated  by a simple mathematical property possessed
by multi-partite states, unlike EPR pairs, which forces them to behave
differently  when measured  in computational  or diagonal  basis. 
Let $H$,  $\bigoplus$ and $\bigotimes$  denote the Hadamard  gate, the
XOR  operation and the  tensor product  respectively then  (with the
presence of a proper normalizing factor in each expression),
\begin{eqnarray}
H^{\bigotimes    n}   |1\rangle^{\bigotimes    n} &=&
\sum_{x_1,x_2,\cdots,x_n}   (-1)^{x_1 + x_2 +   \cdots
+ x_n} |x_1 x_2 \cdots x_n\rangle \nonumber \\
H^{\bigotimes n}|0\rangle^{\bigotimes n} &=& \sum_{x_1,x_2,\cdots,x_n}
|x_1 x_2 \cdots x_n\rangle  \nonumber 
\end{eqnarray}

\begin{eqnarray}
 \therefore H^{\bigotimes n} (|1\rangle^{\bigotimes  n} + |0\rangle^{\bigotimes n}
) &=&  \sum_{x_1\bigoplus x_2\bigoplus \cdots \bigoplus x_n  = 0} |x_1
x_2   \cdots   x_n\rangle   \nonumber   \\   &=&   (\sum_{x_1\bigoplus
x_2\bigoplus \cdots  \bigoplus x_s =  0} |x_1 x_2 \cdots  x_s\rangle )
(\sum_{x_{s+1} \bigoplus  x_{s+2} \bigoplus \cdots \bigoplus  x_n = 0}
|x_{s+1}    x_{s+2}   \cdots   x_n\rangle    )   \nonumber    \\   &+&
(\sum_{x_1\bigoplus x_2\bigoplus  \cdots \bigoplus  x_s = 1}  |x_1 x_2
\cdots x_s\rangle )  (\sum_{x_{s+1} \bigoplus x_{s+2} \bigoplus \cdots
\bigoplus x_n = 1} |x_{s+1} x_{s+2} \cdots x_n\rangle ) \nonumber
\end{eqnarray}

We can observe by symmetry that the above factoring can be infact done
for any  two groups of sizes  $s$ and $n-s$ respectively.   We are now
ready to develop our protocol which involves the following steps:
\begin{description}
\item{(1)} EPR protocol: Along the $n-1$ edges of the minimum spanning
EPR tree, EPR  pairs are created.  This involves  pairwise quantum and
classical communication between any  two parties connected by an edge.
Successful completion  ensures that each  of the two parties  across a
given    edge     share    a    nearly     perfect    singlet    state
$\frac{1}{\sqrt{2}}(|01\rangle - |10\rangle)$.  At the end of the run,
let the  minimum number of EPR  pairs distilled along any  edge of the
minimum spanning EPR tree be $2m$.
\item{(2)} The $2m$ instances of  the singlet state are then converted
to the triplet state $\frac{1}{\sqrt{2}}(|00\rangle + |11\rangle)$, by
the  Pauli operator  $XZ$ being  applied by  the second  party (called
${\cal Y}$) on his qubit.
\item{(3)} For each edge, the  party ${\cal Y}$ intimates the protocol
leader (say  ``Lucy'') of the completion  of step (2). Lucy  is the one
who  starts  and  directs   the  following  protocol  for  teleporting
entanglement. Lucy can be from any of the two groups.
\item{(4)}  Entanglement teleportation  protocol:  Using purely  local
operations and classical communication (LOCC), the $n$ parties execute
the teleportation protocol of \cite{skp}, which consumes the  $n-1$ EPR pairs  to produce one
instance of  the $n$-GHZ state state Eq.  (\ref{maxeq}) shared amongst
them.
\item{(5)}  A projective measurement  in the  {\it diagonal}  basis is
performed by all the parties on their respective qubits.
\item{(6)} Lucy decides randomly a set of $m$ bits to be used as check
bits, and announces their positions.
\item{(7)} All parties from a group assist to get one cbit
corresponding to each check bit position by XORing their corresponding
check bits.  This  gives an effective check bit  corresponding to each
check bit  position. The  two group then  announce the value  of their
effective check bits. If too few of these values agree, they abort the
protocol.  We can  note from the mathematics developed  above that the
effective   check  bits   should  agree   after  the   diagonal  basis
measurement.  Effective non-check bits are also calculated as above by
XORing the non-check bits of the goup members.
\item{(8)} Error correction is done as in for quantum key distribution
between two trustful parites.
\end{description}

{\em  Proof  of unconditional  security (Sketch).~}  The  crucial element  that
simplifies the proof of the above  protocol is that it can potentially be
reducible to
the proof of  the security of sharing bipartite  entanglement. This is
because,  beyond  step  (1),   only  local  operations  and  classical
communication are involved.  We can  thus exclude the involvement of a
malicious eavesdropper Eve beyond  step (1). Apart from correcting for
quantum and classical noise,  and the availability of an authenticated
classical channel, the ability to  detect Eve in this step suffices to
secure the protocol against Eve.

The problem of  secure bipartite entanglement generation has
been extensively studied.  For example, we may assume that step (1) is
implemented  using  the  Lo-Chau  \cite{lochau}  or  Modified  Lo-Chau
protocol  \cite{sp00} (by  leaving  out the  final measurement  step),
which have been proven to be unconditionally secure.

Further,  we  need to  mention  the  role  of fault  tolerant  quantum
computation and of quantum error correcting codes during the execution
of  entanglement teleportation  protocol of  Ref.  \cite{skp}  for the
following reason:  suppose the probability of  error on a  bit is $p$.
Then the probability of an error on the bit obtained by XORing all the
$s$   group  members'   bits   may   be  larger,   given   by:  $P   =
\sum_{r=1,3,5...}   C(s,r)  p^r (1-p)^{s-r}$,  where  $C(s,r)$ is  the
number of  all possible way selecting  $r$ elements from a  set of $s$
distinct elements.  If  $P$ is too close to  $0.5$, then the effective
channel capacity  $Ch$ for  the protocol  (given by $Ch  = 1  - H(P)$,
where $H(.)$  is Shannon entropy) will almost  vanish.  Therefore, the
quantum part  of the  protocol implementation should  be very  good to
ensure that $P$ is not too close to $0.5$.

Of the  XORed $2m$ raw  bits, $m$ bits  are first used for  getting an
estimate of  $P$, by obtaining  the Hamming distance  $\delta$ between
each group's $m$-bit check string.   If they are mutually too distant,
the protocol  run is aborted.  If $\delta$ is  not too great  (that is
$2\delta  + 1  \geq  d$), it  can  be corrected  with  a classical  code
$C(m,k,d)$, where $m$ is block  length, $d$ is (minimum) code distance
and $k/m$ is code rate.   Each group decodes its XOR-ed $m$-bit string
to  the  nearest  codeword  in  $C(m,k,d)$.  This  $k$-bit  string  is
guaranteed  with high  probability  to be  identical  between the  two
groups. An binary enumeration of the $2^k$ codewords of $C(m,k,d)$ can
be used as the actual key shared between the two groups.

Recalling the fact that the security of the QKD requires 
the two involved parties to be honest,  
one important point which should be carefully considered is that
whether the above reduction of the security proof to the bipartite case
is impervious to the attacks when some subset of parties from one group
collude with that from another group. 
It seems that the step 7 can ensure that the worst they can do is 
to sabotage the protocol, however we feel that a more careful study
is required beforing making any such claim and 
we plan to do this in our future studies.      

\acknowledgments SKS  thanks LAMP Group, Raman  Research Institute and
NMR Research Center(SIF),  Indian Institute  of Science  for supporting  his  visit, during
which a part of this work was done.

\end{document}